\documentclass[aps,pre,twocolumn,showpacs]{revtex4}

\usepackage{graphicx}
\usepackage{amsmath}
\usepackage{amssymb}
\usepackage{enumerate}
\usepackage{times}

\DeclareMathAlphabet{\bi}{OML}{cmm}{b}{it}
\DeclareMathOperator{\atan}{tan^{-1}}
\newcommand{\va}{{\bf a}}
\newcommand{\vb}{{\bf b}}

\newcommand{\ve}{{\bf e}}
\newcommand{\vn}{{\bf n}}
\newcommand{\vp}{{\bf p}}
\newcommand{\vx}{{\bf x}}
\newcommand{\vrr}{{\bf r}}

\begin{document}

\title{Fundamental-measure density functional for the fluid
of aligned hard hexagons: New insights in fundamental measure theory}

\author{Jos\'e A.~Capit\'an}
\email{jcapitan@math.uc3m.es}
\author{Jos\'e A.~Cuesta}
\email{cuesta@math.uc3m.es}
\affiliation{Grupo Interdisciplinar de Sistemas Complejos (GISC),
Departamento de Matem\'aticas, Universidad Carlos III de Madrid,
Avenida de la Universidad 30, E--28911 Legan\'es, Madrid, Spain}

\begin{abstract}
In this article we obtain a fundamental measure functional for the
model of aligned hard hexagons in the plane. Our aim is not just to
provide a functional for a new, admittedly academic, model, but
to investigate the structure of fundamental measure theory. A model
of aligned hard hexagons has similarities with the hard disk model.
Both share ``lost cases'', i.e.\ admit configurations of three
particles in which there is pairwise overlap but not triple overlap.
These configurations are known to be problematic for fundamental measure
functionals, which are not able to capture their contribution correctly.
This failure lies in the inability of these functionals to yield a
correct low density limit of the third order direct correlation function.
Here we derive the functional by projecting aligned hard cubes on the
plane $x+y+z=0$. The correct dimensional crossover behavior of these
functionals permits us to follow this strategy. The functional of
aligned hard cubes, however, does not have lost cases, so neither had
the resulting functional for aligned hard hexagons. The latter exhibits,
in fact, a peculiar structure as compared to the one for hard disks.
It depends on a uniparametric family of weighted densities through a
new term not appearing in the functional for hard disks. Apart from
studying the freezing of this system, we discuss the implications of the
functional structure for new developments of fundamental measure
theory.
\end{abstract}


\maketitle

\section{Introduction}

Fundamental measure (FM) theory \cite{rosenfeld:1989} is one of the most
successful density functional (DF) theories, yet also one of the most difficult
to adapt to new systems. Unlike classical DF approximations \cite{evans:1992},
which describe a general approximate recipe in which some knowledge of
the fluid is ``cooked up'' to produce a functional, FM theory constructs functionals
from geometric principles in a far more involved manner. As a result of this
process the resulting functionals have got nicer features; among them, three
are striking: their higher predictive power (they yield structure functions that
are needed as input in classical approximations), their natural formulation
for multicomponent mixtures, and their good behavior under dimensional crossover
(when $d$-dimensional systems are constrained to $d-1$ dimensions, $d$-dimensional
FM functionals become $(d-1)$-dimensional ones).
However, the price to pay for having such nice functionals is that
their structure is extremely rigid: almost any reasonable modification one
makes to ``improve'' the quality of the results spoils one of the above features,
mainly dimensional crossover \cite{tarazona:2002,roth:2002,cuesta:2002}. The
latter is not only a remarkable property that FM functionals (and only them)
share with exact ones, but also a desirable property of any functional which is
meant to study fluids under strong confinement
\cite{figueroa:2003,gonzalez:2006,goulding:2001}.

This is the reason why every extension of FM theory beyond the hard spheres
fluid for which it was originally proposed \cite{rosenfeld:1989,kierlik:1990,
phan:1993} has become a ``major achievement''. Extensions are nowadays available
for parallel hard cubes and parallelepipeds \cite{cuesta:1996,cuesta:1997a,
cuesta:1997b,martinez-raton:1999} (which provide a restricted orientation model
of liquid crystals \cite{martinez-raton:2004}); soft spherical potentials
\cite{schmidt:1999}; non-additive mixtures \cite{schmidt:2000,schmidt:2001a,
schmidt:2004b}; mixtures of rods, plates and spheres \cite{schmidt:2001b,
brader:2002,esztermann:2004,harnau:2005,esztermann:2006}; lattice fluids
\cite{lafuente:2002a,lafuente:2002b,lafuente:2003a,lafuente:2003b,lafuente:2004,
lafuente:2005,cuesta:2005}; and fluids in porous media \cite{schmidt:2002a,
schmidt:2003b,reich:2004}.
Even for hard spheres, Rosenfeld's original functional \cite{rosenfeld:1989}
has undergone important improvements over the years \cite{rosenfeld:1996,
rosenfeld:1997,tarazona:1997,tarazona:2000}, after realizing that dimensional
crossover was a unique feature of this type of functionals very much entangled
to its construction procedure.

When Rosenfeld first conceived FM theory \cite{rosenfeld:1989} it rested
strongly on geometrical properties of spherical overlaps and on scaled-particle
theory \cite{reis:1959}. A decade later, the theory had been reformulated in
terms of ``zero-dimensional (0D) cavities'' \cite{rosenfeld:1996,rosenfeld:1997}.
By this it must be understood a cavity able to hold no more than a hard sphere.
Under the requirement that confinement of the FM functional to one such cavity must
lead to the exact result, and introducing one- two- and three-point cavities,
by adding and subtracting the necessary terms so as to maintain the exact
0D limit a functional arises with the required structure
\cite{tarazona:1997,tarazona:2000}. The result is not ``perfect'' in the
sense that there are three-point 0D cavities for which the
exact result cannot be recovered: those for which three spheres placed at the
three points of the cavity have pairwise overlap but not triple overlap.
These cavities were termed ``lost cases'' \cite{tarazona:1997} because they
do not contribute to the free energy, and their existence reveals the inability
of this construction to reproduce the lowest order in the density expansion
of the three-particle direct correlation function (DCF) \cite{cuesta:2002}
(which is non-zero for those configurations).
As a matter of fact, the problems arising from the extension of this functional
to mixtures of hard spheres have the same origin, and corrections trying to
palliate these problems are unable to remedy the defect of the correlations
\cite{cuesta:2002}.

On the contrary, the FM functional for aligned hard
parallelepipeds does not have lost cases because for this kind of particles,
whenever three particles have pairwise overlap there is necessarily triple
overlap. In fact, the exact 0D limit is recovered for
cavities of any shape \cite{cuesta:1997a}, and the low-density limit of the
three-particle DCF is exact. This led to the
belief that FM theory is simply unable to produce a functional for hard
spheres without lost cases \cite{cuesta:2002}.

\begin{figure}
\includegraphics[width=50mm]{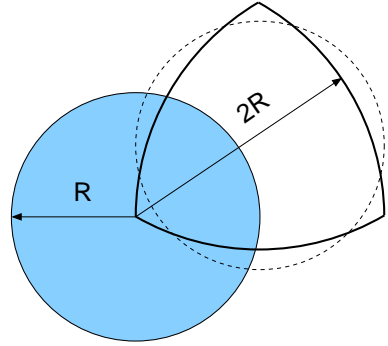}
\caption[]{(color online) Example of a non-spherical maximal 0D cavity for the system
of two-dimensional hard spheres (disks) of radius $R$ (colored circle in the
figure). Any rotation of this cavity will produce a new maximal 0D cavity.
To illustrate that this cavity is not contained in a spherical maximal 0D
cavity, the latter is plotted on top with dashed line.}
\label{fig:max0D}
\end{figure}

The extension of FM theory to lattice fluids is based on this
0D cavity reformulation \cite{lafuente:2002a,lafuente:2002b,
lafuente:2003a,lafuente:2003b,lafuente:2004,lafuente:2005,cuesta:2005}.
But because lattice geometry lacks spherical symmetry, this extension has
uncovered an important ingredient in the theory. While the
FM functional for hard spheres is expressed in terms of a set of weighted
densities whose weights are associated to geometrical features of the
\emph{particles} \cite{rosenfeld:1989,kierlik:1990,phan:1993,rosenfeld:1996,
rosenfeld:1997,tarazona:1997,tarazona:2000}, the weights in lattice FM
functionals are associated to geometrical features of \emph{maximal
0D cavities} \cite{lafuente:2002b,lafuente:2004,lafuente:2005}.
These are 0D cavities such that if they get extended in any way, 
they are not 0D cavities anymore \footnote{Although it is of
little concern here, it is worthwhile to notice that the
definition of 0D cavity introduced in \cite{lafuente:2005}
extends that of \cite{lafuente:2002b} to include cases in which particle
interactions are soft. Thus, according to this new definition, a 0D
cavity is a cavity in which, if there is more than one particle, at least
two of them interact.}. Maximal 0D cavities need not have the same shape
as the particles that define them (in most cases they will not); for
instance, maximal 0D cavities of a nearest-neighbor exclusion lattice
gas in a triangular lattice, which is represented by hard hexagons,
are equilateral triangles (see Ref.~\cite{lafuente:2003b} for this and
other examples). At the same time, by construction these lattice FM
functionals recover the exact 0D limit for \emph{any} 0D cavity 
\cite{lafuente:2004,lafuente:2005} and as a consequence of this fact,
they can be proven to yield the correct low-density limit of the
three-particle DCF \cite{lafuente:2005}. Interestingly, maximal 0D cavities
for aligned hard parallelepipeds have exactly the same shape as the particles,
however this is not true for spheres, where apart from spherical cavities there
are other maximal 0D cavities with different shapes (see Fig.~\ref{fig:max0D}
for an example).

\begin{figure}
\includegraphics[width=50mm]{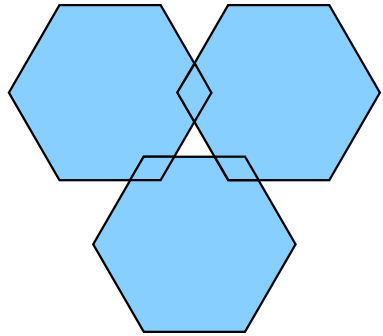}
\caption[]{(color online) Configuration of three aligned hexagons exhibiting pairwise,
but not triple overlap (lost cases).}
\label{fig:lostcaseshex}
\end{figure}

In Ref.~\cite{lafuente:2003b} FM functionals for many two- and
three-dimensional hard core lattice gases were obtained from the known
functional of hard (hyper)cubes in a (hyper)cubic lattice by exploiting
dimensional crossover. One of these was the hard hexagons model, which
was obtained from the hard cubes model by constraining the centers of
mass of the cubes to lay on the plane $x+y+z=0$. In this paper we will
apply the same procedure to the continuum FM functional of parallel hard
cubes in order to obtain the (continuum) two-dimensional fluid of aligned
hard hexagons.
The reason to do this is the following. Three aligned hard hexagons can be
arranged in such a way that there is pairwise overlap but no triple
overlap (see Fig.~\ref{fig:lostcaseshex});
hence, according to the cavity construction of the FM functional
for hard spheres \cite{tarazona:1997,tarazona:2000}, there should be ``lost
cases''. However, we are going to obtain such a functional by dimensional
crossover from the functional of parallel hard cubes which \emph{does not}
have lost cases. As explained, this means that the 0D limit is recovered
for any 0D cavity; therefore the same will hold for the resulting functional
for aligned hard hexagons.

The FM functional for aligned hard hexagons that we will obtain provides
interesting insights into FM theory. First of all, the maximal 0D cavities,
not the particles, are the relevant constructive geometrical object. Secondly,
it points out that a FM functional for hard spheres (or disks) will probably
have an infinite number of terms. From a practical point of view this is good
and bad news: good, because we know what the FM functional for hard spheres
\cite{tarazona:2000} is missing in order to get rid of the lost cases; bad
because a functional with an infinite number of terms will be useless for
real purposes. At the end of this paper we will discuss these issues in more
depth. We think, however, that despite its eventual utility, the information
that this FM functional for aligned hard hexagons provides is relevant for a
thorough understanding of FM theory.

The rest of the paper is organized as follows. In section~\ref{sec:cubestohexagons}
we describe the construction of the FM functional for aligned hard hexagons
by dimensional crossover of the FM functional for the fluid of parallel hard cubes.
The procedure as well as the resulting weighted densities and the form of the
functional are explained in this section, but the detailed calculations
are deferred to Appendix~\ref{sec:appendix}. In Sec.~\ref{sec:thermodynamics}
we analyze the equation of state derived from the functional both for the
uniform fluid and for the triangular solid phase. The functional predicts a
first order freezing at coexisting packing fractions $\eta_f=0.58$ for the
fluid and $\eta_s=0.63$ for the solid. Finally, in
section~\ref{sec:discussion} we discuss the features of the resulting functional,
with special emphasis on those that cause the functional to be free from lost
cases.

\section{From hard cubes to hard hexagons}
\label{sec:cubestohexagons}

\begin{figure}
\includegraphics[width=60mm]{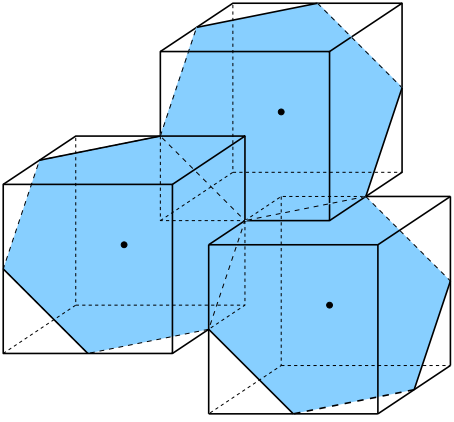}
\caption[]{(color online) Constraining the centers of mass of parallel hard cubes
to lay on the plane $x+y+z=0$ yields a system of aligned hard
hexagons.}
\label{fig:cubesection}
\end{figure}

As described in Ref.~\cite{lafuente:2003b} the way to obtain an
effective system of aligned hard hexagons is to start off from a
system of hard cubes aligned parallel to the coordinate axes and
constraint their centers of mass to lay on the plane $x+y+z=0$.
Figure~\ref{fig:cubesection} illustrates this geometrical construction.
Making use of the good behavior of FM functionals under any dimensional
crossover, we will carry out the same projection in the FM functional
of parallel hard cubes and thus obtain the one for aligned hexagons.

The FM functional of parallel hard cubes
of edge-length $2L$ can be written as
\cite{cuesta:1997a,cuesta:1997b}
\begin{equation}
F_\mathrm{PHC}[\tilde\rho]=
F_\mathrm{PHC}^\mathrm{id}[\tilde\rho]+F_\mathrm{PHC}^\mathrm{ex}[\tilde\rho],
\label{eq:FPHC}
\end{equation}
where
\begin{equation}
\beta F_\mathrm{PHC}^\mathrm{id}[\tilde\rho]=\int \mathrm{d}\vrr\,
\tilde\rho(\vrr)(\ln\mathcal{V}\tilde\rho(\vrr)-1)
\end{equation}
is the ideal contribution ($\mathcal{V}$ is the thermal volume and
$\beta=(kT)^{-1}$, with $k$ the Boltzmann constant and $T$ the temperature)
for a density profile of the hard cube fluid $\tilde\rho(\vrr)$, and
\begin{equation}
\beta F_\mathrm{PHC}^\mathrm{ex}[\tilde\rho]=\int \mathrm{d}\vrr\,
\Phi_\mathrm{PHC}\big(\{p_{\alpha}(\vrr)\}\big),
\label{eq:excessPHC}
\end{equation}
with
\begin{eqnarray}
\Phi_\mathrm{PHC} &=& \Phi_\mathrm{PHC}^{(1)}+ \Phi_\mathrm{PHC}^{(2)}+
\Phi_\mathrm{PHC}^{(3)}, \\
\Phi_\mathrm{PHC}^{(1)} &=& -\frac{1}{8}p_0\ln(1-p_3),\label{eq:firstPHC}\\
\Phi_\mathrm{PHC}^{(2)} &=& \frac{\vp_1\cdot\vp_2}{8(1-p_3)},\\
\Phi_\mathrm{PHC}^{(3)} &=& \frac{p_{2,1}p_{2,2}p_{2,3}}{8(1-p_3)^2},
\end{eqnarray}
is the excess, over the ideal, free energy. The functions $p_{\alpha}(\vrr)$
are weighted densities
\begin{equation}
p_{\alpha}(\vrr)=\int \mathrm{d}\vrr'\,\omega_{\alpha}(\vrr-\vrr')
\tilde\rho(\vrr'),
\label{eq:wdensPHC}
\end{equation}
where the scalar or vectorial weights are given by
\begin{eqnarray}
\omega_3(\vrr) &=& \tau(x)\tau(y)\tau(z), \label{eq:weights1}\\
{\boldsymbol\omega}_2(\vrr) &=& \big(\zeta(x)\tau(y)\tau(z),
\tau(x)\zeta(y)\tau(z), \nonumber \\ && \tau(x)\tau(y)\zeta(z)\big),\\
{\boldsymbol\omega}_1(\vrr) &=& \big(\tau(x)\zeta(y)\zeta(z),
\zeta(x)\tau(y)\zeta(z), \nonumber \\ && \zeta(x)\zeta(y)\tau(z)\big),\\
\omega_0(\vrr) &=& \zeta(x)\zeta(y)\zeta(z) \label{eq:weights4},
\end{eqnarray}
with
\begin{equation}
\tau(u)=\Theta(L-|u|), \qquad
\zeta(u)=\delta(L-|u|),
\end{equation}
$\Theta(x)$ being Heaviside's step function ($0$ if $x<0$ and $1$ if $x>0$) and
$\delta(x)$ Dirac's delta. Notice that $p_{2,j}$ ($j=1,2,3$) denotes the $j$th
component of $\vp_2$.

\begin{figure}
\includegraphics[width=40mm]{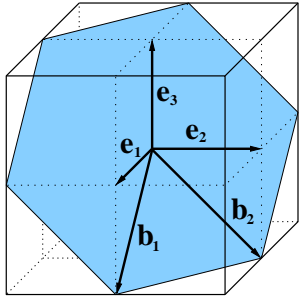}
\caption[]{(color online) Choice of an appropriate basis on the plane $x+y+z=0$ to represent
the coordinates of the hexagons.}
\label{fig:basis}
\end{figure}

Now, the projection amounts to taking
\begin{equation}
\tilde\rho(\vrr)=\rho(\vx)\delta(x+y+z)
\label{eq:projection}
\end{equation}
in the functional (\ref{eq:FPHC}), where $\rho(\vx)=\rho(x,y)$ is the
density profile of aligned hard hexagons. The choice of coordinates corresponds
to a change to the (non-orthogonal) basis $\{\vb_1,\vb_2,\vb_3\}$
given by $(\vb_1,\vb_2,\vb_3)=(\ve_1,\ve_2,\ve_3)\,P$, with
\begin{equation}
P =\begin{pmatrix} \phantom{-}1 & \phantom{-}0 & \phantom{-}0 \\
\phantom{-}0 & \phantom{-}1 & \phantom{-}0 \\ -1 & -1 & \phantom{-}1 \end{pmatrix}
\label{eq:matrix}
\end{equation}
and $\{\ve_1,\ve_2,\ve_3\}$ the canonical basis. Vectors $\vb_1$ and $\vb_2$
form a basis on the plane $x+y+z=0$ (see Fig.~\ref{fig:basis}). This choice of
vectors amounts to working with the projections of the hexagons on the XY
plane, because the projections of $\vb_1$ and $\vb_2$ are simply $\ve_1$ and
$\ve_2$.

\begin{figure}
\includegraphics[width=60mm,clip=]{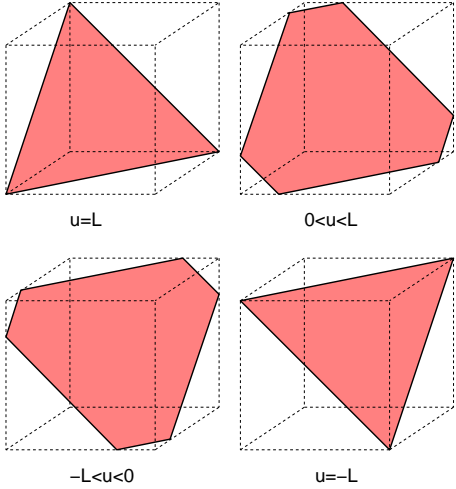}
\caption[]{(color online) Maximal 0D cavities for the system of aligned hard
hexagons are obtained as the sections of the cubes by planes
$x+y+z+u=0$ with $-L\le u\le L$. Triangular cavities correspond
to $u=\pm L$ while hexagonal ones to $-L<u<L$ ($u=0$ is the 
regular hexagon).}
\label{fig:hexcavities}
\end{figure}

The details of introducing the density profile (\ref{eq:projection}) into
the excess part of the free-energy functional are deferred to
Appendix~\ref{sec:appendix}. Here we simply give the final result.
The projection transforms the original weighted densities
for the cubes, $p_{\alpha}(\vrr)$, into a set of new densities for
the hexagons. The most striking result is that these new weighted
density are associated to maximal 0D cavities for the hexagons, not
to the hexagons themselves. The complete set of such maximal 0D
cavities can be obtained as the sections of one of the original
cubes by the planes $x+y+z+u=0$, where $-L\le u\le L$ (see
Fig.~\ref{fig:hexcavities}). The cases $u=\pm L$ correspond to
two equilateral triangles (pointing up and down), while the cases $-L<u<L$
correspond to hexagons (of which only $u=0$ is a
regular hexagon identical to the fluid particles). This comes as
an important difference with respect to the FM functional for hard
spheres, and in retrospect yields a new interpretation to the
weighted densities of the original cubes as associated to maximal
0D cavities (which in the case of cubes are indistinguishable from the
particles).

\begin{figure}
\includegraphics[width=60mm,clip=]{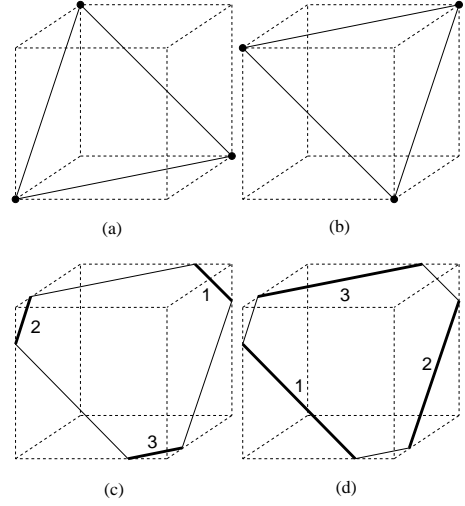}
\caption[]{Weighted densities are associated to geometric elements
of the cavities: $n_0^{(+)}(\vx)$ (a) and $n_0^{(-)}(\vx)$ (b) to the vertices
of the triangular cavities; the components of $\vn_1^{(+)}(\vx,u)$
(c) and $\vn_1^{(-)}(\vx,u)$ (d) to the edges of the cavities. In
(c) and (d) the numbers 1, 2, 3 label the component which is associated
to each edge.}
\label{fig:densities}
\end{figure}

To be precise, the weighted densities that we need to express the
functional are
\begin{eqnarray}
n_0^{(\pm)}(\vx) &=& \int\mathrm{d}\vx'\,\Omega_0^{(\pm)}
(\vx-\vx')\rho(\vx'), \label{eq:n0}\\
\vn_1^{(\pm)}(\vx,u) &=& \int\mathrm{d}\vx'\,
\boldsymbol\Omega_1^{(\pm)}(\vx-\vx',u)\rho(\vx'), \label{eq:n1}\\
n_2(\vx,u) &=& \int\mathrm{d}\vx'\,\Omega_2(\vx-\vx',u)\rho(\vx'),
\label{eq:n2}
\end{eqnarray}
and it is convenient to define also the weighted densities
\begin{eqnarray}
\vn_1^{(\pm)}(\vx) &=& \vn_1^{(\pm)}(\vx,\pm L), \label{eq:n1L}\\
n_2^{(\pm)}(\vx) &=& n_2(\vx,\pm L) \label{eq:n2L}.
\end{eqnarray}
The weights that define these densities are the following:
\begin{eqnarray}
\Omega_0^{(\pm)}(\vx) &=& \delta(x\mp L)\delta(y\mp L) \nonumber\\
&&+\delta(x\mp L)\delta(y\pm L) \\
&&+\delta(x\pm L)\delta(y\mp L), \nonumber \\
\boldsymbol\Omega_1^{(\pm)}(\vx,u) &=&
\Big(\delta(x\mp L)\tau(y)\tau(u-x-y), \nonumber\\
&& \tau(x)\delta(y\mp L)\tau(u-x-y), \\
&& \tau(x)\tau(y)\delta(u-x-y\mp L)\Big), \nonumber \\
\Omega_2(\vx,u) &=& \tau(x)\tau(y)\tau(u-x-y).
\end{eqnarray}

The meaning of these weighted densities is related to averages over
different geometric elements of the maximal 0D cavities to which
they are associated. Thus, $n_2(\vx,u)$ is the average over the
area of the cavity corresponding to that value of $u$ (the colored
regions in Fig.~\ref{fig:hexcavities}); each component of 
$\vn_1^{(\pm)}(\vx,u)$ is the average over one edge of the hexagonal
cavity (triangular if $u=\pm L$); and $n_0^{(\pm)}(\vx)$ is the
average over the three vertices of the corresponding triangular cavity.
The two latter cases are illustrated in Fig.~\ref{fig:densities}.

With the help of these weighted densities we can write
$\beta F_\mathrm{AHH}^\mathrm{ex}[\rho]=\int\mathrm{d}\vx\,\Phi_\mathrm{AHH}(\vx)$,
where
\begin{eqnarray}
\Phi_\mathrm{AHH} &=& \Phi_\mathrm{AHH}^{(1)}+ \Phi_\mathrm{AHH}^{(2)}+
\Phi_\mathrm{AHH}^{(3)}, \label{eq:onetwothree} \\
\Phi_\mathrm{AHH}^{(1)} &=& -\frac{1}{6}\sum_{\pm}n_0^{(\pm)}\ln(1-n_2^{(\pm)}),\\
\Phi_\mathrm{AHH}^{(2)} &=& \frac{1}{6}\sum_{\pm}\frac{n_{1,1}^{(\pm)}n_{1,2}^{(\pm)}
      +n_{1,2}^{(\pm)}n_{1,3}^{(\pm)}+n_{1,3}^{(\pm)}n_{1,1}^{(\pm)}}
      {1-n_2^{(\pm)}}, \nonumber \\
&& \label{eq:second}\\
\Phi_\mathrm{AHH}^{(3)} &=& \frac{1}{2}\sum_{\pm}\int_{-L}^L\mathrm{d}u\,
      \frac{n_{1,1}^{(\pm)}(u)n_{1,2}^{(\pm)}(u)n_{1,3}^{(\pm)}(u)}{[1-
      n_2(u)]^2}. \nonumber \\
&& \label{eq:third}
\end{eqnarray}
(For the sake of notational simplicity we have omitted the argument $\vx$ in
all weighted densities, retaining only the argument $u$ in those that depend
on it.)

There are several features worth noticing in this FM functional for aligned
hard hexagons which we have derived from the one for parallel hard cubes.
First of all, the most obvious fact: weighted densities are associated to
the geometry of maximal 0D cavities, as in lattice FM functionals, and not 
to the geometry of particles, as in the FM functional for hard spheres or
disks. Secondly, as in the system of aligned hard hexagons there is an
infinity of maximal 0D cavities, the third term exhibits a ``sum'' 
over them all; hence the integral in that term. Finally, the typical FM
structure for $\Phi$ as a sum of $D$ terms,
$D$ being the dimension of the problem, breaks down here: we have a
\emph{two}-dimensional system which is described as a sum of \emph{three}
terms.

All these features will have consequences for the general structure of
FM functionals, which we shall discuss later in Sec.~\ref{sec:discussion}

\section{Thermodynamics of the fluid of aligned hard hexagons}
\label{sec:thermodynamics}

\subsection{Fluid phase}

The free energy of the fluid phase is obtained by specializing the
weighted densities with a uniform particle density. The only subtle
point we have to take into account is that, because of our choice of
coordinates (which actually describe the projections of hexagons on
the XY plane) if $\rho$ denotes the particle density of hexagons (measured
on the plane $x+y+z=0$), the uniform density profile will reduce to
$\rho(\vx)=\sqrt{3}\rho$. (This $\sqrt{3}$ is the scale factor difference
between actual hexagons and their projections.) With this in mind, and
given that the area of a hexagon is $v_h=3\sqrt{3}L^2$ (recall that $L$ is
half the edge length of the cubes), hence the packing fraction 
$\eta=v_h\rho$, the weighted densities reduce in this limit to
\begin{eqnarray}
n_0 &=& \eta/L^2, \\
\vn_1^{(\pm)}(u) &=& \frac{\eta}{3L}\left(1\pm\frac{u}{L}\right)\,(1,1,1),\\
\vn_1^{(\pm)} &=& \frac{2\eta}{3L}\,(1,1,1),\\
n_2(u) &=& \left(1-\frac{u^2}{3L^2}\right)\eta, \\
n_2^{(\pm)} &=& \frac{2\eta}{3},
\end{eqnarray}
where the packing fraction $\eta=v_h\rho$,
and therefore the excess free energy per unit volume (in $kT$ units), $\Phi$,
becomes
\begin{equation}
\begin{split}
\frac{\Phi}{\rho}=&\,-\ln\left(1-\frac{2\eta}{3}\right)+\frac{\eta}{3(1-\eta)}
+\left(3-\frac{8}{3}\eta\right) \\
&\times\sqrt{\frac{\eta}{3(1-\eta)^3}}
\atan\sqrt{\frac{\eta}{3(1-\eta)}}.
\end{split}
\end{equation}

{}From this expression we can obtain the equation of state as
\begin{equation}
\begin{split}
\frac{\beta p}{\rho} &= 1+\eta\frac{\partial(\Phi/\rho)}{\partial\eta} \\
&= \frac{1-\eta/2}{(1-\eta)^2}+\left(\frac{3}{2}-\eta\right) \\
&\times \sqrt{\frac{\eta}{3(1-\eta)^5}}\atan\sqrt{\frac{\eta}{3(1-\eta)}}.
\end{split}
\label{eq:eos}
\end{equation}
Perhaps the most striking feature of this equation of state is its divergence
as $(1-\eta)^{-5/2}$ at close packing. This exponent $2.5$ is noticeably higher
than the exponent $2$ that a straightforward scaled particle argument would predict.


\subsection{Solid phase}

The standard way to approach the solid phase in DF theory is to use
a parametrization of the density profile as a sum of Gaussians centered at
the nodes of the solid lattice,
\begin{equation}
\hat\rho(\vx')=\frac{\theta\alpha}{\pi}\sum_{r_1,r_2\in\mathbb{Z}}
\exp\big\{-\alpha(\vx'-rd\va'_1-sd\va'_2)^2\big\}.
\label{eq:rhosol}
\end{equation}
Here  $\hat\rho(\vx')$ denotes the density profile of hexagons on the plane
$x+y+z=0$; $\vx'=(x',y')$ denotes the position referred to an orthogonal coordinate
system on that plane; $\alpha$ is related to the mean square displacement of
the particle with respect to its lattice node; $\theta$ is the occupancy of
the solid (the mean number of particles in one unit cell), which accounts for
the vacancies; and $d=d_\mathrm{c}\sqrt{\theta/\eta}$ is the lattice parameter,
with $\eta$ the packing fraction and $d_\mathrm{c}$ the lattice parameter at
close packing, $d_\mathrm{c}=\sqrt{6}L$ (hexagons have edge length $\sqrt{2}L$).
The unit vectors $\va'_1=(0,1)$ and $\va'_2=(-\sqrt{3}/2,1/2)$ are a convenient
choice for the basis of the triangular lattice's unit cell.

\begin{figure}
\includegraphics[width=60mm,clip=]{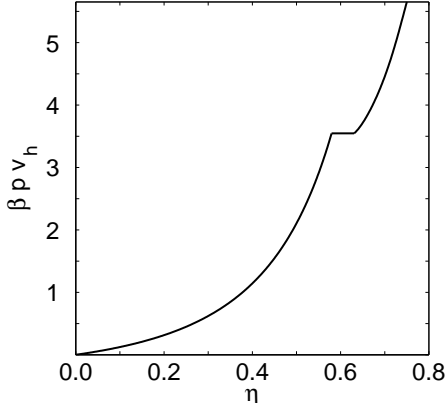}
\caption[]{Reduced pressure, $\beta pv_h$, with $v_h=3\sqrt{3}L^2$ the
area of a hexagon, versus pacing fraction $\eta=\rho v_h$ for a fluid of
aligned hard hexagons. The packing fractions at coexistence are $\eta_f=
0.58$ for the fluid and $\eta_s=0.63$ for the solid.}
\label{fig:eossolid}
\end{figure}

Although the implementation of this density profile is rather straightforward,
a few words on the appropriate choice of variables may be helpful. First of
all, as all weighted densities are expressed in terms of projected coordinates
on the plane XY, we should describe the density profile in terms of these
coordinates. Thus we can write  $\vx'=J\vx$, with
\begin{equation}
J =\begin{pmatrix} \sqrt{2} & \sqrt{1/2} \\
\phantom{-}0 & \sqrt{3/2} \end{pmatrix},
\end{equation}
and, given that $\det J=\sqrt{3}$, the density profile becomes
$\rho(\vx)=\sqrt{3}\hat\rho(\vx')$.
In the projected representation, the unit cell is a rhombus, so it is convenient
to introduce in the integrals the change of variables $\vx \rightarrow Q\vx$,
where
\begin{equation}
Q =\frac{1}{\sqrt{6}}\begin{pmatrix} -1 & -2 \\
\phantom{-}2 & \phantom{-}1 \end{pmatrix}.
\end{equation}
This transforms the global integral as
\begin{equation}
\int\limits_{\mbox{\scriptsize unit cell}}\hspace*{-3mm}\mathrm{d}\vx
\longrightarrow \frac{1}{2}
\int_{-d/2}^{d/2}\mathrm{d}x\int_{-d/2}^{d/2}\mathrm{d}y
\end{equation}
(notice that $\det Q=1/2$).

In the projected coordinates, the sum of Gaussians defining the density
profile can be factorized as
\begin{equation}
\rho(\vx)=\sqrt{3}\theta\sum_{r_1,r_2\in\mathbb{Z}}
g_{\alpha}\left(\frac{x_{r_1}}{2}+y_{r_2}\right)
g_{\alpha}\Big(\frac{\sqrt{3}\,x_{r_1}}{2}\Big),
\end{equation}
with the definitions $x_{r_1}=x-r_1d$, $y_{r_2}=y-r_2d$ and
$g_{\alpha}(x)=\sqrt{\alpha/\pi}\,\mathrm{e}^{-\alpha x^2}$. 
This permits one to express the weighted densities as products of
Gaussian and error functions.

\begin{figure}
\includegraphics[width=60mm,clip=]{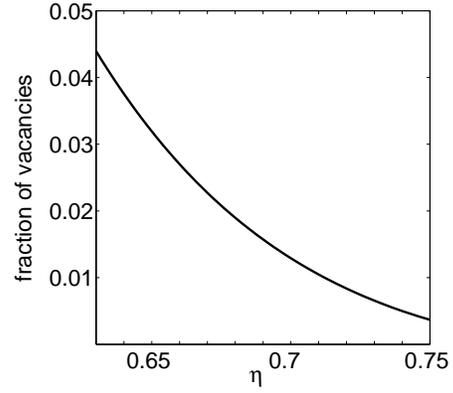}
\caption[]{Fraction of vacancies, $1-\theta$, as a function of the
packing fraction $\eta$ for the fluid phase.}
\label{fig:vacancies}
\end{figure}

\begin{figure}
\includegraphics[width=60mm,clip=]{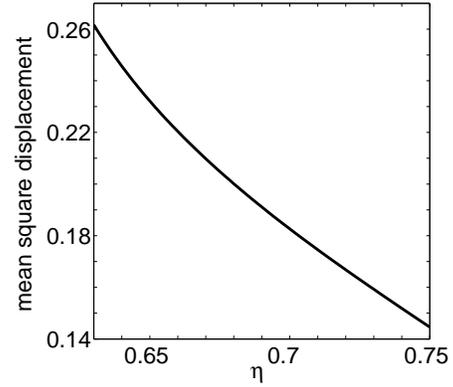}
\caption[]{Mean square displacement of particles around the equilibrium
positions in the solid phase, as a function of the packing fraction $\eta$.}
\label{fig:msd}
\end{figure}

Minimization of the functional is carried out numerically in the parameters
$\alpha$ and $\theta$. This determines the free energy of the solid phase for
every packing fraction $\eta$. At the point where this free-energy curve
branches off the one of the fluid phase, the slope discontinuously decreases.
This produces a concavity in the whole free-energy curve and therefore the
transition is first order and the coexisting densities are determined via a
standard Maxwell's double-tangent construction. The resulting equation of state,
depicting both the fluid and the solid pressures, is plotted in
Fig.~\ref{fig:eossolid}. Also plotted in Figs.~\ref{fig:vacancies} and
\ref{fig:msd} are the fraction of vacancies, $1-\theta$, and the square
root of the mean square displacement of hexagons with respect to their
lattice positions. Interestingly, the solid of hard hexagons
has a low fraction of vacancies all the
way up from the transition (never larger than 5\%), in marked contrast
with what happens for the fluid of hard cubes from which the functional
for this system is derived \cite{martinez-raton:1999}.

One last remark is in order. The result found here for the equation of
state of the system of aligned hard hexagons is very similar to that of
its lattice counterpart \cite{lafuente:2003b}. For the latter the exact
result is known to have a continuous transition \cite{baxter:1982}, although
so smooth that a first order discontinuity is not a bad quantitative
approximation. The exact result for the continuum model is unknown,
but certainly the same caveat on the nature of the transition applies to it.

\section{Discussion}
\label{sec:discussion}

The fluid of aligned hard hexagons has strong similarities with the
fluid of hard disks. The most important for the aims of this work is
that particles may be placed in configurations such that they have
pairwise overlap but not triple overlap (see Fig.~\ref{fig:lostcaseshex}).
These configurations have been termed ``lost cases'' \cite{tarazona:1997}
because FM theory, as currently formulated for hard spheres or hard
disks \cite{tarazona:1997,tarazona:2000}, is unable to capture their
contribution. The basic constructive principle of FM theory is the
recovery of the exact 0D limit of the free energy when the system is
constraint to any 0D cavity. Lost cases arise in certain 0D cavities
(for instance, for hard disks, a circular cavity of radius larger than
$R$ and smaller than $2R/\sqrt{3}$) and thus the FM functional does
not recover the exact limit for them. This failure of the theory is
associated to an incorrect low density limit of the third order
direct correlation function \cite{cuesta:2002}; in other words,
the density expansion of the FM functional for hard disks or spheres
is incorrect already at third order.

The logic of the construction of FM functional for $d$-dimensional
hard spheres requires that the excess free energy density be a
sum of $d$ terms \cite{tarazona:1997}
\begin{equation}
\Phi_{d{\rm-HS}}=\Phi_{d{\rm-HS}}^{(1)}+
\Phi_{d{\rm-HS}}^{(2)}+\cdots+\Phi_{d{\rm-HS}}^{(d)}.
\end{equation}
Further terms constructed on the same logic are identically zero.
Each of these terms is incorporated starting from the first one
and trying to compensate for the spurious terms that arise when
two, three, etc., particles are incorporated to a 0D cavity. When
there are lost cases, the last one vanishes and thus cannot bring
about its compensation. The logic of this construction strongly
relies on the fact that weights have the same shape of the particles,
as in Rosenfeld's original FM theory \cite{rosenfeld:1989}, of which
this new constructive method is just a generalization. 

Applying the same logic to the fluid of aligned hard hexagons would
lead to the same result and lost cases would arise. Yet, there is another
method to obtain the FM functional for such a fluid, which is the
projection we have carried out in this work of the fluid of parallel
hard cubes on a specific plane. The latter fluid does not have lost
cases because of the particular form of its particles (if there
is pairwise overlap between three cubes, there is necessarily
triple overlap as well), and this nice property is inherited by
the functional for the hexagons. As a matter of fact, the resulting
functional has a different structure is several respects. First of
all, there is a weighted density for every one of the maximal 0D
cavities conceivable for this system. These include two triangles
and a continuum of irregular hexagons. Because of this, weighted
densities depend on a parameter that gauges the shape of the cavity.
This feature is not new: it was revealed in the study of FM functionals
for lattice models \cite{lafuente:2002b,lafuente:2003a,lafuente:2003b,%
lafuente:2004,lafuente:2005}. But second and more importantly, there
appears a new term, say a ``$d+1$ term'', thanks to which the compensation
to recover the exact 0D limit for any 0D cavity is guaranteed. As
explained above, this term cannot be predicted by applying the
construction logic of the hard disk functional.

Actually both features are connected: there is an extra term because
there are several 0D maximal cavities that contribute weighted
densities to the functional. What this is telling us is that the
functional for hard disks or hard spheres is simply incomplete.
Cavities such as that shown in Fig.~\ref{fig:max0D} (and possibly other)
should make their contribution through new weighted densities. Notice
in passing that any rotation of the cavity of the figure is a new,
different, 0D cavity and so there should also be a continuum of weighted
densities, as for the hexagons. And accordingly, new terms beyond
$\Phi_{d{\rm-HS}}$ are to be expected. How many of them? We do not
have a definitive answer to this question, but we will provide convincing
arguments that there will be infinitely many.

The projection we have carried out from three-dimensional cubes to
aligned hard hexagons can be generalized to obtain a FM functional
for $2l$-gons in a straightforward manner. For instance, projections
of the fourth-dimensional system of hard hypercubes on the appropriate
plane generates octagons. In general, projecting $l$-dimensional
hypercubes on an appropriate plane generates aligned hard $2l$-gons.
Because of the structure of the fluid of parallel hard hypercubes
\cite{cuesta:1997a} we know that such functionals will have $l$
terms. Disks are obtained as the limit $l\to\infty$ of $2l$-gons,
so in this limit, the FM functional will be an infinite series.
It might happen that the series can be summed up and produce just
a single (more complex) term, but this can hardly be foreseen.

So, should we seek for a FM functional for hard disks or hard
spheres without lost cases? Well, from the arguments above we
believe that it would be a pointless task, for even if we could
overcome the difficulty of its construction, we would probably
end up with an extremely cumbersome functional, useless for
practical purposes. This does not mean that perhaps including
some, not all, the missing terms we could obtain improvements
on the current functional. This point might be worth exploring 
in the future.

\acknowledgments

We acknowledge very useful discussions with Luis Lafuente and
Pedro Tarazona. This work is funded by projects MOSAICO, from Ministerio de
Educaci\'on y Ciencia (Spain),
MOSSNOHO, from Comunidad Aut\'onoma de Madrid (Spain), and
CCG06-UC3M/ESP-0668, from Universidad Carlos
III de Madrid and Comunidad Aut\'onoma de Madrid (Spain).
The first author also acknowledges financial support through a
contract from Consejer\'{\i}a de Educaci\'on of Comunidad de Madrid 
and Fondo Social Europeo.

\appendix
\section{Projection of the FM functional for parallel hard cubes on the
plane $\boldsymbol{x+y+z=0}$}
\label{sec:appendix}

\subsection{Dimensional crossover}

Using the projection on the XY plane defined by $\vrr \rightarrow P\vrr$,
with $P$ given by (\ref{eq:matrix}),
\begin{equation}
\beta F^\mathrm{ex}_\mathrm{PHC}[\tilde\rho]=
\int \mathrm{d}\vrr\, \Phi_\mathrm{PHC}(\{p_{\alpha}(P\vrr)\}),
\end{equation}
where
\begin{equation}
p_{\alpha}(P\vrr)=\int \mathrm{d}\vrr'\,\omega_{\alpha}(P(\vrr-\vrr'))
\tilde\rho(P\vrr'),
\end{equation}
with the weights defined in (\ref{eq:weights1})--(\ref{eq:weights4}).
According to (\ref{eq:projection}), in the projected coordinates
\begin{equation}
\tilde\rho(P\vrr)=\rho(\vx)\delta(z).
\end{equation}
This transformation in the density profile allows us to integrate in $z'$, so
the weighted densities become
\begin{equation}
p_{\alpha}(P\vrr)=\int \mathrm{d}\vx'\,\omega_{\alpha}(P(\vx-\vx',z))\rho(\vx'),
\end{equation}
where we are expressing a three-component vector as $\vrr=(\vx,z)$. From
the previous equation one can define a new set of weighted densities
$\{n_{\alpha}\}$, which depend on a parameter $u$ that corresponds to the $z$
coordinate of the above expression,
\begin{equation}
n_{\alpha}(\vx,u)\equiv\int \mathrm{d}\vx'\,\Omega_{\alpha}(\vx-\vx',u)\rho(\vx'),
\end{equation}
with $\alpha=0,1,2$ and $\Omega_{\alpha}$ a scalar or vector function given by
\begin{eqnarray}
\Omega_2(\vx,u) &=& \omega_3(P(\vx,u)),\\
{\boldsymbol\Omega}_1(\vx,u) &=& {\boldsymbol\omega}_2(P(\vx,u)),\\
{\boldsymbol\Omega}_0(\vx,u) &=& {\boldsymbol\omega}_1(P(\vx,u)),
\label{eq:n0vec}\\
\Omega_0(\vx,u) &=& \omega_0(P(\vx,u)).\\ \nonumber
\end{eqnarray}

The next step is to obtain the resulting excess free energy after the
projection. In order to do that, unnecessary degrees of freedom must
be eliminated, which in this particular case amounts to integrating the $z$
coordinate in the PHC functional. Consider the first term (\ref{eq:firstPHC}).
Here the integration is immediate, because $\Omega_0$ is sum of Dirac's
deltas \footnote{In each one of the eight products of three deltas that 
define $\Omega_0(\vx,u)$, the dependence in $u=z$ can be used to evaluate
the integral in the functional. But two of them (those with $u=\pm 3L$) do
not contribute because $n_2(\vx,\pm 3L)=0$. The remaining six terms are
the ones that define $n_0^{(\pm)}$ [see Figs.~\ref{fig:densities} (a) and
(b)].}. Thus we obtain
\begin{equation}
\Phi_\mathrm{AHH}^{(1)}=-\frac{1}{8}\sum_{\pm} n_0^{(\pm)}\ln\left(
1-n_2^{(\pm)}\right),
\end{equation}
with $n_0^{(\pm)}$ and $n_2^{(\pm)}$ defined by (\ref{eq:n0}) and (\ref{eq:n2}),
respectively.

For the second and third terms in (\ref{eq:excessPHC}), direct substitution of
$n_{\alpha}$ leads to
\begin{eqnarray}
\Phi_\mathrm{AHH}^{(2)} &=& \frac{1}{8}\int \mathrm{d}u\,
\frac{\vn_0(u)\cdot\vn_1(u)}{1-n_2(u)},
\label{eq:Phi2u}\\
\Phi_\mathrm{AHH}^{(3)} &=& \frac{1}{8}\int \mathrm{d}u\,
\frac{n_{1,1}(u) n_{1,2}(u) n_{1,3}(u)}{[1-n_2(u)]^2}.
\label{eq:Phi3u}
\end{eqnarray}
Here $n_{\alpha}(u)$ is a shorthand for $n_{\alpha}(\vx,u)$. Notice that the
integration limits in the above formulas are determined by the support of the
weighted density $n_2$, which is a product of Heaviside's step functions. It
is easy to check that this support is $|u|\leq 3L$. Hence, according to the
expression for $n_2(u)$, maximal 0D cavities are recovered (see
Fig.~\ref{fig:hexcavities}) when $|u|\leq L$, but when $L\leq|u|\leq 3L$ the
corresponding cavities are not maximal. The appearance of these non-maximal
0D cavities is a result of our projection procedure, but the functional cannot
explicitly depend on them because any information they provided is, by
definition, already accounted for by the maximal 0D cavities. In what follows
we will explain how to get rid of these spurious contributions.

\subsection{Elimination of spurious terms}

We will show here how to eliminate non-maximal 0D cavities through integration
by parts. To this purpose we introduce the following notation: the weight
$\boldsymbol{\Omega}_1$ can be decomposed as follows (in what follows we will
omit the dependence on $(\vx,u)$ for simplicity, introducing only a dependence
on $u$ whenever ambiguity might arise)
\begin{equation}
\boldsymbol{\Omega}_1=\boldsymbol{\Omega}_1^{(+)}+
\boldsymbol{\Omega}_1^{(-)},
\end{equation}
where $\boldsymbol{\Omega}_1^{(\pm)}$ has the same expression as 
$\boldsymbol{\Omega}_1$ in terms of functions $\tau$ y $\zeta$, but replacing
$\zeta(x)$ with $\delta(x \mp L)$, respectively. New weighted densities
$\vn_1^{(\pm)}$ can also be introduced associated to these weights. These
densities have the properties that $\vn_1^{(+)}$ vanishes when $-3L\le u
\le -L$, whereas $\vn_1^{(-)}$ does it when $L\le u\le 3L$. 

Let us introduce now the differential operator
\begin{equation}
{\bf D}\equiv\left(\frac{\partial}{\partial x}+\frac{\partial}{\partial u},
\frac{\partial}{\partial y}+\frac{\partial}{\partial u},
\frac{\partial}{\partial u}\right).
\end{equation}
Acting on $n_2$ results in
\begin{equation}
{\bf D} n_2=-\vn_1^{(+)}+\vn_1^{(-)}.
\label{eq:n1pm}
\end{equation}
As of now, we will only consider the term $\Phi_\mathrm{AHH}^{(3)}$ 
for $-3L\le u\le -L$, which will be denoted $\Phi_\mathrm{AHH}^{(3,1)}$.
According to (\ref{eq:n1pm}), in this interval the integrand of
$\Phi_\mathrm{AHH}^{(3,1)}$ can be written as
\begin{equation}
\begin{split}
\Phi'''_0(&n_2)n_{1,1}^{(-)}n_{1,2}^{(-)}n_{1,3}^{(-)} =\\
&\frac{1}{3}\sum_{\sigma \in \Pi_3^{+}} \left(D_{\sigma(1)}\Phi''_0(n_2)\right)
n_{1,\sigma(2)}^{(-)}n_{1,\sigma(3)}^{(-)},
\end{split}
\end{equation}
where $\Pi_3^{+}$ is the set of all the permutations $\sigma$ of $\{1,2,3\}$ with
positive signature. Using this identity we can integrate by parts
to obtain
\begin{equation}
\begin{split}
\Phi&_\mathrm{AHH}^{(3,1)} = \frac{n_{1,1}^{(-)}n_{1,2}^{(-)}+
n_{1,2}^{(-)}n_{1,3}^{(-)}+n_{1,3}^{(-)}n_{1,1}^{(-)}}{24[1-n_2^{(-)}]}\\
&- \frac{1}{24}\int_{-3L}^{-L} \!\!\mathrm{d}u\,\Phi''_0(n_2)\!\!
\sum_{\sigma \in \Pi_3^{+}}\!\!D_{\sigma(1)}\!
\left(n_{1,\sigma(2)}^{(-)}n_{1,\sigma(3)}^{(-)}\right)\!(u).
\label{eq:Phi31}
\end{split}
\end{equation}
Notice that in the first term, the components of $\vn_1^{(-)}$ without
explicit dependence on $u$ refer to those of $\vn_1^{(-)}(\vx,-L)$, as in
(\ref{eq:n1L}). The second term in the
expression above can be simplified taking into account the new identity
\begin{equation}
D_{\sigma(2)}n_{1,\sigma(3)}^{(-)}+D_{\sigma(3)}n_{1,\sigma(2)}^{(-)}=
2n_{0,\sigma(1)}.
\label{eq:Dn1n0}
\end{equation}
Recall that $\vn_0$ is defined through the weight (\ref{eq:n0vec}). Besides,
\begin{equation}
\begin{split}
&\sum_{\sigma \in \Pi_3^{+}} D_{\sigma(1)}\left(n_{1,\sigma(2)}^{(\pm)}
n_{1,\sigma(3)}^{(\pm)}\right) = \\
&\sum_{\sigma \in \Pi_3^{+}} n_{1,\sigma(1)}^{(\pm)}
\left(D_{\sigma(2)}n_{1,\sigma(3)}^{(\pm)}+
D_{\sigma(3)}n_{1,\sigma(2)}^{(\pm)}\right),
\label{eq:sumDn1n1}
\end{split}
\end{equation}
so substitution of (\ref{eq:Dn1n0}) and (\ref{eq:sumDn1n1}) into (\ref{eq:Phi31})
allows us to get
\begin{equation}
\begin{split}
\Phi_\mathrm{AHH}^{(3,1)} =& \frac{n_{1,1}^{(-)}n_{1,2}^{(-)}+
n_{1,2}^{(-)}n_{1,3}^{(-)}+n_{1,3}^{(-)}n_{1,1}^{(-)}}{24[1-n_2^{(-)}]}\\
&-\frac{1}{12}\int_{-3L}^{-L} \mathrm{d}u\,\Phi''_0(n_2)\vn_0(u)\cdot\vn_1^{(-)}(u).
\end{split}
\end{equation}
Moreover, as ${\bf D}\cdot \vn_0=0$ and $\Phi''_0(n_2)\vn_1^{(-)}={\bf D}\Phi'_0(n_2)$,
a new integration by parts in the integral above yields
\begin{equation}
\begin{split}
\Phi_\mathrm{AHH}^{(3,1)} =& \frac{n_{1,1}^{(-)}n_{1,2}^{(-)}+
n_{1,2}^{(-)}n_{1,3}^{(-)}+n_{1,3}^{(-)}n_{1,1}^{(-)}}{24[1-n_2^{(-)}]}\\
&+\frac{1}{12}n_0^{(-)}\ln(1-n_2^{(-)}).
\\&
\end{split}
\end{equation}

By symmetry, an identical formula can be obtained when $L\le u\le 3L$,
but with the weighted densities $n_{\alpha}^{(+)}$ instead of $n_{\alpha}^{(-)}$.
Doing exactly the same partial integration in (\ref{eq:Phi2u}) for any
$L\leq|u|\leq 3L$ and gathering all the contributions, we
arrive at the result
\begin{equation}
\begin{split}
\Phi_\mathrm{AHH} =& \,\,\,\Phi_\mathrm{AHH}^\mathrm{res} 
-\frac{1}{6}\sum_{\pm}n_0^{(\pm)}\ln(1-n_2^{(\pm)})+\\
&\,\,\,\frac{1}{24}\sum_{\pm}\frac{n_{1,1}^{(\pm)}n_{1,2}^{(\pm)}+
n_{1,2}^{(\pm)}n_{1,3}^{(\pm)}+n_{1,3}^{(\pm)}n_{1,1}^{(\pm)}}{1-n_2^{(\pm)}},
\label{eq:Phimed}
\end{split}
\end{equation}
where the residual contribution $\Phi_\mathrm{AHH}^\mathrm{res}$ is given by
\begin{equation}
\begin{split}
\Phi_\mathrm{AHH}^\mathrm{res} =& \,\,\,\frac{1}{8}\int_{-L}^L \mathrm{d}u\,
\frac{\vn_0(u)\cdot\vn_1(u)}{1-n_2(u)}+\\
&\,\,\,\frac{1}{8}\int _{-L}^L\mathrm{d}u\,
\frac{n_{1,1}(u) n_{1,2}(u) n_{1,3}(u)}{[1-n_2(u)]^2}.
\label{eq:Phires}
\end{split}
\end{equation}
Because of the range $-L\le u\le L$ of these integrals, Eq.~(\ref{eq:Phimed})
only contains contributions from maximal 0D cavities, as required.

A third integration by parts in the residual term will allow us to write
the functional in a more compact way. According to (\ref{eq:n1pm}), whenever
$|u|\leq L$ we can write 
\begin{equation}
\vn_1=2\vn_1^{(\pm)}\pm {\bf D}n_2.
\end{equation}
Then the last integral in (\ref{eq:Phires}) can be expressed in a symmetric
form combining $\vn_1^{(\pm)}$,
\begin{equation}
\frac{1}{16}\int _{-L}^L\mathrm{d}u\,\Phi'''_0(n_2)
\sum_{\pm}\prod_{j=1}^3\left(2n_{1,j}^{(\pm)}\pm D_j n_2\right)(u).
\end{equation}
We now expand the product above and realize three things. First, that the
product containing three derivatives vanishes. Second, that cross terms
containing one or two derivatives can be integrated with the aid of
(\ref{eq:n1pm}) and
\begin{equation}
D_{\sigma(2)}n_{1,\sigma(3)}^{(\pm)}+D_{\sigma(3)}n_{1,\sigma(2)}^{(\pm)}=
\pm n_{0,\sigma(1)},
\end{equation}
which holds for all $|u|\leq L$. These terms cancel the first one of
(\ref{eq:Phires}) and generate a boundary term identical to the last one in
(\ref{eq:Phimed}). This operation justifies the numerical coefficient appearing
in (\ref{eq:second}). Third, the remaining term, which contains products of
the three components of $\vn_1^{(\pm)}$, produces (\ref{eq:third}). This
way we arrive at the final form of the functional
(\ref{eq:onetwothree})--(\ref{eq:third}).

\newpage
\bibliography{../../Bibliografia/liquid}

\end{document}